\documentclass{article} 

\usepackage{iclr2026_conference,times}

\usepackage{amsmath,amsfonts,bm}

\def\eqref#1{equation~\ref{#1}}

\def\1{\bm{1}}

\DeclareMathAlphabet{\mathsfit}{\encodingdefault}{\sfdefault}{m}{sl}
\SetMathAlphabet{\mathsfit}{bold}{\encodingdefault}{\sfdefault}{bx}{n}

\usepackage{pgfplots}
\pgfplotsset{compat=1.18}
\usepackage[utf8]{inputenc}
\usepackage{amsmath, amssymb}    
\usepackage{graphicx}            
\usepackage{booktabs}            
\usepackage{hyperref}            
\usepackage{longtable}
\usepackage{url}
\usepackage{xurl}
\usepackage{subcaption}
\usepackage{float}               
\usetikzlibrary{positioning, quotes, arrows.meta, fit}
\usepackage[inkscapelatex=false]{svg}
\usepackage{wrapfig}
\usepackage{comment}

\title{
Detecting Information Channels in Congressional Trading via Temporal Graph Learning
}

\author{\textbf{Benjamin Pham Roodman}\thanks{Equal contribution} \\
  National Taiwan University \\
  \texttt{\footnotesize R12922210@ntu.edu.tw} \\
  \And
  \textbf{Eugene Sy}\footnotemark[1] \\
  National Taiwan University \\
  \texttt{\footnotesize d13949006@ntu.edu.tw} \\
  \And
  \textbf{J. Xavier Atero Vázquez} \\
  National Taiwan University \\
  \texttt{\footnotesize d11949007@ntu.edu.tw} \\
  \AND 
  \textbf{Yu-Shiang Huang} \\
  National Taiwan University \\
  and Academia Sinica \\
  \texttt{\footnotesize f09946004@ntu.edu.tw} \\
  \And
  \textbf{Che Lin} \\
  National Taiwan University \\
  \phantom{and Academia Sinica} \\ 
  \texttt{\footnotesize chelin@ntu.edu.tw} \\
  \And
  \textbf{Chuan-Ju Wang} \\
  Academia Sinica \\
  \phantom{and Academia Sinica} \\ 
  \texttt{\footnotesize cjwang@citi.sinica.edu.tw} 
}
\iclrfinalcopy

\begin{document}

\maketitle
\lhead{Under review at ICLR 2026 Workshop on Advances in Financial AI}

\begin{abstract}
Congressional stock trading has raised concerns about potential information asymmetries and conflicts of interest in financial markets. 
We introduce a temporal graph network (TGN) framework to identify information channels through which members of Congress may obtain advantageous knowledge when trading company stocks. 
We construct a multimodal dynamic graph integrating diverse publicly available datasets, concerning congressional stock transactions, lobbying relationships, campaign finance contributions, and geographical connections between legislators and corporations.
Our approach formulates the detection problem as a dynamic edge classification task, where we identify trades that exhibit statistically significant outperformance relative to the S\&P 500 across long time horizons.
To handle the temporal nature of these relationships, we develop a two-step walk-forward validation architecture that respects information availability constraints and prevents look-ahead bias. 
We evaluate several labeling strategies based on risk-adjusted returns and demonstrate that the TGN successfully captures complex temporal dependencies between congressional-corporate interactions and subsequent trading performance. 
\end{abstract}


\section{Introduction}




The integrity of financial markets relies heavily on the transparency of interactions between government officials and corporate entities. Yet, public suspicion that members of Congress leverage privileged information for personal gain has persisted for decades \citep{ziobrowski2004abnormal, karadas2021did}. Following high-profile investigations alleging that legislators traded on non-public intelligence \citep{schweizer2011throw, cbs2011insiders}, the Stop Trading on Congressional Knowledge (STOCK) Act of 2012 was enacted. While intended to affirm insider trading prohibitions and mandate timely financial disclosures \citep{nagy2011insider, belmont2022do}, questions regarding the effectiveness of these regulatory guardrails remain.

Despite these regulations, legislators retain a significant structural advantage over the industries they oversee. This position allows for informational anticipation, where members estimate market-moving events—such as defense contracts or tax code changes—weeks before they become public \citep{williams2024rottenegg}. Such foresight is often facilitated by specific committee assignments, which provide early access to actionable intelligence \citep{hanousek2022dilemma}. The market has increasingly recognized this persistent edge, evidenced by the emergence of investment vehicles that utilize STOCK Act disclosures as a signal. Exchange Traded Funds (ETFs) such as NANC and KRUZ now explicitly track the portfolios of sitting congresspeople, effectively institutionalizing the pursuit of congressional alpha \citep{tidal2024spotlight}.


However, treating these trades merely as isolated time-series signals ignores the rich structural context in which they occur—such as the specific committee ties or lobbying relationships that often precipitate a transaction. While modeling this ecosystem as a dynamic graph offers a promising solution, applying standard Temporal Graph Networks (TGNs) to the financial domain introduces a critical challenge: \textit{information staleness}. Unlike social networks where feedback is often immediate, financial markets exhibit a substantial resolution latency between the observation of a trade and the materialization of its performance label (e.g., 18-month returns). Standard models, which require immediate supervision, typically discard interactions during this unresolved gap. This forces the model to rely on outdated node representations at prediction time, discarding the most recent, and potentially most informative, structural signals.

To address this, we propose \textbf{GAP-TGN} (\textit{\textbf{G}ated \textbf{A}synchronous \textbf{P}ropagation} TGN). 
By formalizing the congressional trading ecosystem as a dynamic bipartite graph and introducing a gated fusion mechanism with asynchronous propagation, GAP-TGN maintains a fresh information state by processing interactions even during the label resolution gap. 
We cast the identification of salient trades as a supervised edge classification task within this latency-aware framework.

Our contributions are threefold:

\begin{enumerate} 

    \item We introduce \textbf{Capitol Gains}, an integrated dataset that unifies legislative disclosures, lobbying records, and campaign finance into a coherent dynamic graph structure.
    \item We propose \textbf{GAP-TGN}, a novel temporal graph architecture designed for financial interactions. We introduce a \textit{Gated Multi-Modal Fusion} layer forcombining dynamic memory with static attributes and an \textit{Asynchronous Propagation} strategy that keeps node embeddings fresh even in the absence of immediate labels.
    \item Preliminary experiments demonstrate that GAP-TGN outperforms purely tabular baselines in long-term horizons. While baselines suffer from mode collapse, GAP-TGN maintains robust performance, offering a proof-of-concept that relational position is a leading indicator of political alpha.

\end{enumerate}






Beyond these initial findings, we position this work as a catalyst for future research. We identify the legislative-corporate graph as an under-explored frontier for financial graph learning. 
We conclude by discussing the potential for integrating complex, higher-order interaction types—such as bill sponsorship cycles and lobbying intensity—laying the groundwork to decode the high-dimensional mechanics of political alpha.

\section{Data Construction}

To strictly model the information asymmetry between Capitol Hill and Wall Street, we introduce \textbf{Capitol Gains}, a multimodal temporal graph dataset. 
Unlike prior works that analyze congressional trading in isolation, our goal is to contextualize these financial transactions within a broader network of legislative influence. 
The dataset unifies three distinct domains—legislative activity, corporate lobbying, and financial markets—into a dynamic graph structure evolving daily from January 1, 2013, to December 31, 2025.

\begin{figure}
    \centering
    \vspace{-10pt} 
    \includegraphics[width=0.70\textwidth, trim=0 0 {0.0\textwidth} 0, clip]{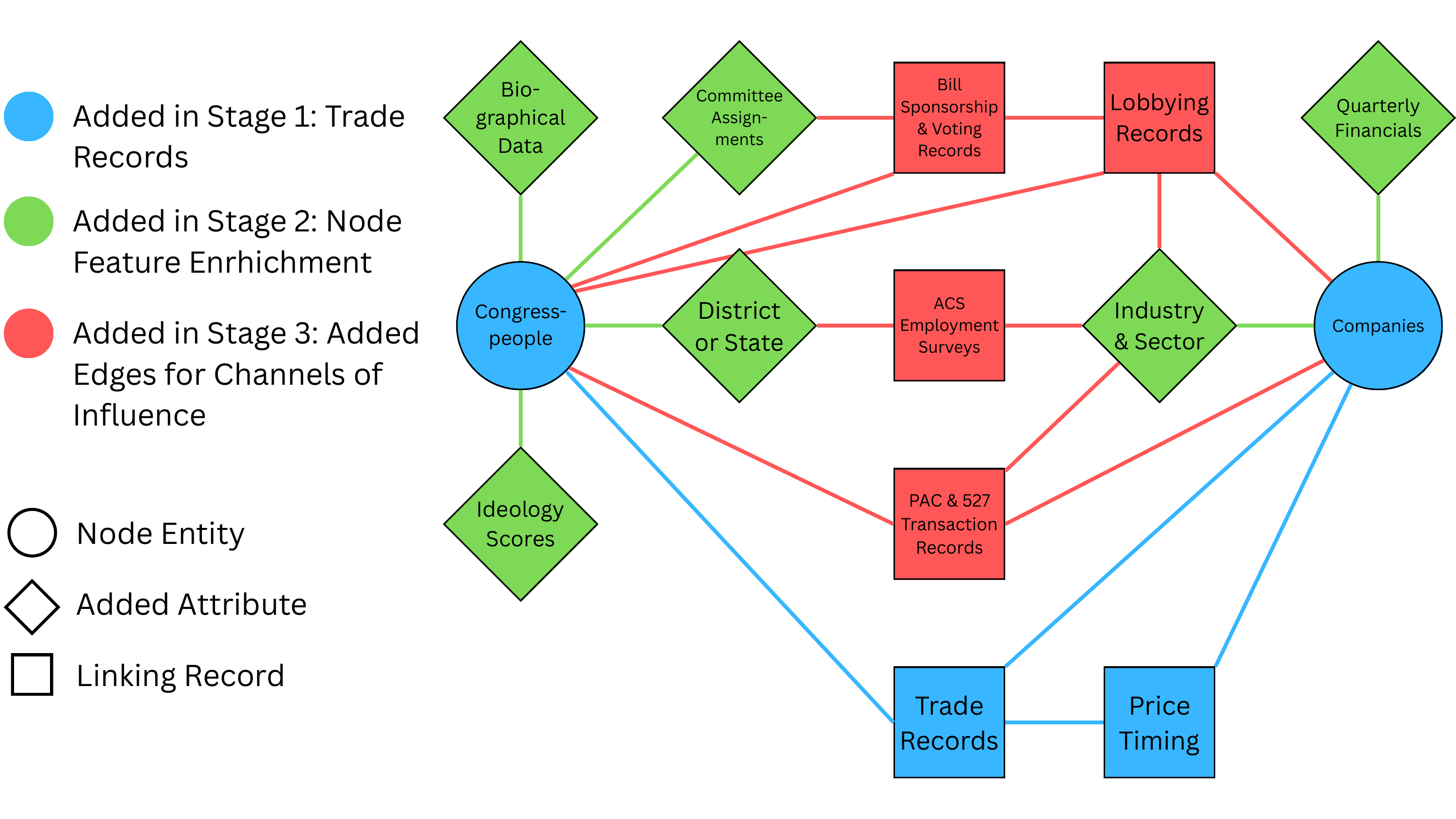}
    \caption{The relational graph of our data.}
    \label{fig:data_schema}
    \vspace{-10pt} 
\end{figure}



A critical design constraint of this dataset is the strict prevention of lookahead bias. We enforce a \textbf{Point-in-Time (PIT)} construction protocol that strictly aligns data availability with public disclosure. 
Under this regime, congressional transactions are integrated into the graph solely upon their filing date, the moment the market becomes aware of the trade---rather than the private execution date. 
We extend this temporal discipline to macroeconomic indicators and corporate fundamentals, indexing all features by their precise release timestamps. 
This synchronization ensures that the model's information state at any given time $t$ corresponds exactly to the information set available to market participants, maintaining strict causality.


The resulting graph schema, visualized in Figure \ref{fig:data_schema}, synthesizes the heterogeneous datasets into two primary node sets, Congresspeople and Companies, connected by the dynamic interactions defined below. 
The statistics of the dataset are detailed in Table \ref{tab:statistics} in Appendix~\ref{app:statistics}



\subsection{Congresspeople Information}
We characterize the legislative nodes using biographical, geographical, and ideological attributes to capture their potential access to private information.

\textbf{Legislative Profiles:} We utilize the \texttt{congress-legislators}\footnote{https://github.com/unitedstates/congress-legislators} repository \citep{congress_legislators} for core biographical data, tenure, and party affiliation. To capture local economic incentives, we map legislators to their districts' industrial makeup using employment data from the \textbf{American Community Survey (ACS)}, applying zip-code crosswalking for years prior to standardized district-level reporting.

\textbf{Dynamic Ideology and Committees:}
Legislators' positions are not static. We track their evolving political alignment using daily W-NOMINATE scores derived from roll-call votes \citep{lewis2021voteview}, utilizing the first two dimensions to represent economic and social positioning (updated with a two-year rolling lookback). Furthermore, committee assignments, often the source of privileged regulatory information, are embedded using a multi-hot encoding scheme based on a fixed vocabulary of all standing and select committees.

\subsection{Transactions and Influence Records}
Having characterized the legislative agents and their attributes, we now delineate the dynamic mechanisms that link them to the corporate sector.
We distinguish between two categories of interactions: the explicit \textbf{financial signal} derived from direct stock transactions, and the broader \textbf{relational context} established through lobbying efforts and campaign contributions. These edges collectively map the flow of both capital and political influence.

\textbf{Congressional Trading Records:}
Our primary target edges are stock transactions sourced from \textbf{Quiver Quantitative} \citep{quiverquant}. We filter the raw data to isolate equity-based signals, removing non-stock assets. Each edge contains the trade direction (Buy/Sell), temporal metadata (lag between trade and file date), and volume. 
Consistent with the STOCK Act, transaction values are reported in buckets (e.g., \$1K--\$15K); we utilize the floor of these buckets for conservative volume estimation.

\textbf{Lobbying and Campaign Finance Records:}
We map two channels of corporate influence. 
First, we integrate \textbf{LobbyView} \citep{kim2018lobbyview} to link corporate clients to specific bills. Using \textbf{VoteView} data, we then connect these bills to legislators. We define a "Strong Connection" edge if a legislator sponsored a lobbied bill, and a "Weak Connection" edge if they merely voted 'Yea' on it.
Second, we aggregate campaign contributions from \textbf{OpenSecrets}, linking corporate Political Action Committees (PACs) and 527 Organizations to legislators. These edges quantify the direct financial support a congressperson receives from the industries they regulate.

\subsection{Financial and Market Data}
We constructed a dataset of daily price and volume indicators. Daily OHLCV (Open, High, Low, Close, Volume) data for all equities was privately sourced from the \textbf{Massive API} (formerly Polygon.io) \citep{massive_data}. 
We augmented the standard price data with Dark Pool Trading Volume, derived by aggregating off-exchange trade reports for each ticker.

\subsection{Economic and Corporate Context}
To capture the broader economic environment, we incorporate 64 macroeconomic indicators sourced from the Federal Reserve Bank of St. Louis' API for Archival Federal Reserve Economic Data (ALFRED) and corporate financial fundamentals from quarterly 10-Q filings retrieved via the U.S. Securities and Exchange Commission's API for Electronic Data Gathering, Analysis, and Retrieval (EDGAR) \citep{sec_edgar, fred_alfred}. These features—detailed in Appendices~\ref{app:macro_indicators} and~\ref{app:corporate_facts} respectively—are updated strictly based on their public release or filing dates rather than the reporting period, ensuring the model's information state reflects only data available to market participants at the time of any given transaction.

\section{Methodology}
\label{sec:methodology}

\subsection{Problem Definition}

We formalize the detection of information channels in congressional trading as a dynamic edge classification task on a continuous-time heterogeneous graph.

\textbf{Graph Formulation.} We model the congressional ecosystem as a continuous-time heterogeneous graph $\mathcal{G}(t) = (\mathcal{V}, \mathcal{E}(t))$. The node set $\mathcal{V} = \mathcal{V}_L \cup \mathcal{V}_C$ partitions the entities into \textbf{Politicians} (Legislators) and \textbf{Companies} (Issuers), each associated with time-varying feature vectors $\mathbf{x}_v(t)$ for $v \in \mathcal{V}$.
The graph evolves through a temporal event stream $\mathcal{S} = \{e_1, e_2, \dots, e_N\}$.
We formally define the \textit{historical context} $\mathcal{H}(t) = \{ e_j \in \mathcal{S} \mid t_j < t \}$ as the set of all preceding events available at time $t$.
Each interaction event involves either two nodes or a single node; those with two nodes are each defined as a tuple $e = (u, v, t, k, \mathbf{x})$, where $u \in \mathcal{V}_L$ and $v \in \mathcal{V}_C$ are the interacting nodes at time $t$, $k \in \mathcal{K}$ denotes the interaction type, and $\mathbf{x}_{k}(t) \in \mathbb{R}^{d_k}$ captures the event attributes. Interaction events that update attributes on a single node are defined as $e = (u,t,x)$, with $u\in \mathcal{V}$ and $t,x$ unchanged. 
In addition to the event streams and node attributes, we incorporate a systematic market feature vector $\mathbf{x}_{\text{mkt}}$.

\textbf{Target Definition.} To structure the learning task, we designate a specific interaction type as the \textit{target relation} ($k_{\text{target}} \in \mathcal{K}$), while treating all other interaction types as \textit{relational context} ($k \in \mathcal{K} \setminus \{k_{\text{target}}\}$). This framework allows the model to leverage the latent influence structure provided by context edges to interpret explicit signals.
Focusing strictly on the case where $k_{\text{target}} = \textsc{Trade}$, consider a target transaction $e_i = (u, v, t, \textsc{Trade}, \mathbf{x})$. We define a binary label $y_i$ based on its excess return over a future horizon $\Delta t$. Letting $R_{v}(t, \Delta t)$ and $\bar{R}(t, \Delta t)$ denote the cumulative asset and benchmark returns respectively, we set:

\vspace{-0.5cm}
\begin{equation}\label{eq:target}
    y_i = \mathbb{I}\left( \left( R_{v}(t, \Delta t) - \bar{R}(t, \Delta t) \right) > \tau \right)
\end{equation}
\vspace{-0.5cm}

where $\tau$ is a performance threshold. The objective is to learn a mapping $f(\mathcal{H}(t), \mathbf{X}(t), e_i) \to \hat{y}_i$ that predicts these high-alpha opportunities given the historical context $\mathcal{H}(t)$ and evolving node states $\mathbf{X}(t)$.
\vspace{-0.5cm}

\subsection{Gated Asynchronous Propagation Temporal Graph Network}

\begin{figure}[htbp]
    \centering
    \includegraphics[width=0.93\textwidth]{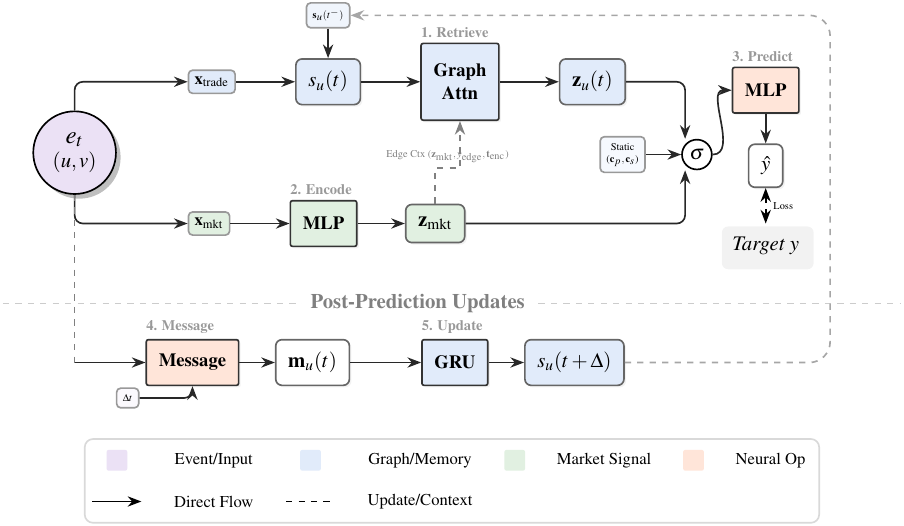}
    \caption{\textbf{GAP-TGN Architecture} Processing an event $e_t(u,v)$ involves: (1) Encoding trade and market features; (2) Retrieving memory $s_u(t)$ via graph attention modulated by edge context; (3) Fusing dynamic embeddings with static attributes ($\mathbf{c}_p, \mathbf{c}_s$) to predict $\hat{y}$; (4) Constructing a message from the event signal and $\Delta t$ post-prediction; and (5) Propagating the update to the memory state $s_u(t+\Delta t)$, strictly separating prediction time from state evolution.}
    \label{fig:gap-tgn-architecture}
    \vspace{-0.4cm}
\end{figure}

We propose \textbf{GAP-TGN} (\textit{\textbf{G}ated \textbf{A}synchronous \textbf{P}ropagation} Temporal Graph Network), which extends the Temporal Graph Network (TGN) architecture \citep{rossi2020temporalgraphnetworksdeep} for Continuous-Time Dynamic Graphs (CTDGs). Unlike standard TGNs designed for high-frequency, immediate-feedback environments (e.g., social networks), GAP-TGN is explicitly designed to address the \textbf{information staleness} caused by long-horizon prediction windows ($\Delta t$) and the multi-modal heterogeneity inherent in the \textit{Capitol Gains} dataset.

\subsubsection{Model Inputs and Feature Encoding (Steps 0--2)}
\label{sec:inputs}
We first define the raw input signals and their latent encodings (Step 2) that drive the downstream components.

\textbf{Feature Inputs.}
Let $\mathbf{x}_{\text{trade}}$ denote the transaction attribute vector and $\mathbf{x}_{\text{mkt}}$ denote the systematic market feature vector (detailed in Sec. \ref{sec:experimental_setup}).
We also define a harmonic time encoding $\mathbf{t}_{\text{enc}}(\Delta t)$ to capture temporal relative positions. Additionally, each legislator node $u$ is associated with learnable static embeddings $\mathbf{c}_p$ (Party) and $\mathbf{c}_s$ (State) to capture time-invariant affiliations.

\textbf{Propagated Supervision Signal} ($y_{\text{edge}}$).
To mitigate information staleness due to the resolution latency $\Delta T$, we introduce a latent state ($y=0.5$) to the label set:
\begin{equation}
    y_{\text{edge}} =
    \begin{cases}
       1 & \text{Outperformed (Resolved)} \\
       0 & \text{Underperformed (Resolved)} \\
       0.5 & \text{Pending (Latent)}
    \end{cases}
\end{equation}

\textbf{Latent Market Encoding (Step 2).}
The raw market features are projected into a latent space via a Multi-Layer Perceptron (MLP) to form the \textit{Market Signal}:
\begin{equation}
    \mathbf{z}_{\text{mkt}}(t) = \texttt{MLP}(\mathbf{x}_{\text{mkt}}(t))
\end{equation}
This latent signal $\mathbf{z}_{\text{mkt}}$ is utilized in two distinct downstream components: as context for historical retrieval (Step 1) and as a global signal for gated fusion (Step 3).

\subsubsection{Temporal Embedding and Context Retrieval (Step 1)}
To classify a target interaction at time $t$, the model first queries the node's current memory state $\mathbf{s}_u(t)$. It retrieves the local neighborhood $\mathcal{N}_u(t)$---defined as the sequence of the $M$ most recent interactions involving node $u$ prior to $t$. We utilize a Graph Attention mechanism to aggregate these historical signals into a temporal embedding $\mathbf{z}_u(t)$:
\vspace{-0.2cm}

\begin{equation}
    \mathbf{z}_u(t) = \text{MultiHeadAttention}\left( \mathbf{q}_u, \mathbf{K}_u, \mathbf{V}_u \right)
\end{equation}
Specifically, for each attention head $h$, we define the query $\mathbf{q}^{(h)}_u$, key $\mathbf{k}^{(h)}_j$, and value $\mathbf{v}^{(h)}_j$ vectors. The query represents the current state of the focal node, while the keys and values encode the historical interaction context:
\begin{align}
    \mathbf{q}^{(h)}_u &= \mathbf{W}_q^{(h)} \mathbf{s}_u(t) \\
    \mathbf{k}^{(h)}_j &= \mathbf{W}_k^{(h)} \mathbf{s}_j(t^-) + \mathbf{W}_{k,e}^{(h)} \mathbf{\phi}_j \\
    \mathbf{v}^{(h)}_j &= \mathbf{W}_v^{(h)} \mathbf{s}_j(t^-) + \mathbf{W}_{v,e}^{(h)} \mathbf{\phi}_j
\end{align}
where $t^-$ denotes the timestamp just before $t$, $\mathbf{W}_q, \mathbf{W}_k, \mathbf{W}_v$ are learnable weight matrices, and $\boldsymbol{\phi}_j$ is the edge feature vector.
We compute the attention coefficients $\alpha_{u,j}^{(h)}$ using a scaled dot-product attention, which determines the relevance of each historical neighbor $j \in \mathcal{N}_u(t)$ to the current prediction task:
\begin{equation}
    \alpha_{u,j}^{(h)} = \text{softmax}_j \left( \frac{(\mathbf{q}^{(h)}_u)^\top \mathbf{k}^{(h)}_j}{\sqrt{d_k}} \right)
\end{equation}
The final embedding $\mathbf{z}_u(t)$ is formed by concatenating the weighted sums from all $H$ heads:
\begin{equation}
    \mathbf{z}_u(t) = \Bigg\Vert_{h=1}^{H} \left( \sum_{j \in \mathcal{N}_u(t)} \alpha_{u,j}^{(h)} \mathbf{v}^{(h)}_j \right)
\end{equation}
This mechanism enables the model to effectively query its historical context. For instance, if the current query $\mathbf{q}_u$ encodes a technology sector purchase, the attention weights $\alpha_{u,j}$ will dynamically concentrate on past neighbors where $\boldsymbol{\phi}_j$ indicates similar structural patterns or high-alpha transactions.

The attention scores are conditioned on the \textbf{Edge Context} $\mathbf{\phi}_j$. Unlike the memory update, the attention mechanism explicitly incorporates the outcome labels ($y_{\text{edge}}$) of historical neighbor interactions.

The edge feature vector $\mathbf{\phi}_j$ is composed of the pre-defined inputs:
\begin{equation}
    \boldsymbol{\phi}_j = [\mathbf{x}_{\text{trade}}(t) \parallel \mathbf{z}_{\text{mkt}}(t) \parallel \mathbf{t}_{\text{enc}}(\Delta t) \parallel y_{\text{edge}}]
\end{equation}
The inclusion of $y_{\text{edge}}$ enables the attention head to discriminate between neighbors associated with positive alpha, negative alpha, or pending outcomes.

\subsubsection{Gated Multi-Modal Fusion (Steps 2--3)}
A critical challenge in this domain is distinguishing between idiosyncratic insider alpha and systematic market momentum.
To address this, the GAP-TGN architecture incorporates a \textbf{Gated Fusion} layer (Step 3) that adaptively weights structural and market signals. We define two distinct latent representations:

\textbf{Structural Signal} ($\mathbf{z}_{\text{graph}}$): The concatenation of the temporal graph embedding and learnable static embeddings representing the legislator's Party and State affiliation:
\[ \mathbf{z}_{\text{graph}} = [\mathbf{z}_u(t) \parallel \mathbf{c}_{p} \parallel \mathbf{c}_{s}] \]

\textbf{Market Signal} ($\mathbf{z}_{\text{mkt}}$): As defined in Section \ref{sec:inputs}, this is the latent encoding of the technical indicators.

A learnable scalar gate $g \in (0,1)$ arbitrates the contribution of each modality:
\begin{equation}
    g = \sigma\left( \mathbf{W}_g [\mathbf{z}_{\text{graph}} \parallel \mathbf{z}_{\text{mkt}}] + b_g \right)
\end{equation}
\begin{equation}
    \hat{y} = \texttt{MLP}\left( [g \cdot \mathbf{z}_{\text{graph}} \parallel (1-g) \cdot \mathbf{z}_{\text{mkt}}] \right)
\end{equation}

where $\sigma$ denotes the sigmoid activation function.

This formulation allows the model to dynamically shift its decision boundary, prioritizing structural graph patterns (high $g$) when trading behavior deviates from market fundamentals, and relying on momentum indicators (low $g$) when structural signals are weak.

\subsubsection{Memory Dynamics and State Persistence (Steps 4--5)}
After the prediction phase, the model updates the state of the participating nodes to reflect the new interaction. Each node $v \in \mathcal{V}$ maintains a persistent memory state $\mathbf{s}_v(t) \in \mathbb{R}^{d_m}$, serving as a compressed representation of its historical interaction trajectory.

\textbf{Message Computation (Step 4).}
To prevent label leakage during the training process, we enforce a strict separation between the memory update signal and the supervision signal. The message function receives strictly observable interaction attributes defined by $\mathbf{x}_{\text{trade}}$.
The message $\mathbf{m}_u(t)$ is computed systematically:
\begin{equation}
    \mathbf{m}_u(t) = \texttt{MLP}\left( \mathbf{s}_u(t^-) \parallel \mathbf{s}_v(t^-) \parallel \Delta t \parallel \mathbf{x}_{\text{trade}} \right)
\end{equation}
where $\Delta t$ denotes the time elapsed since the previous event.


\textbf{Memory Update (Step 5).}
The node state is subsequently updated via a GRU unit:
\begin{equation}
\mathbf{s}_u(t) = \texttt{GRU}(\mathbf{s}_u(t^-), \mathbf{m}_u(t)).
\end{equation}

\subsubsection{Training Objective}
The model is trained end-to-end to minimize the weighted binary cross-entropy loss:
\begin{equation}
    \mathcal{L} = -\frac{1}{N} \sum_{i=1}^N w_{\text{pos}} \cdot y_i \log(\sigma(\hat{y}_i)) + (1-y_i) \log(1 - \sigma(\hat{y}_i))
\end{equation}
where $w_{\text{pos}}$ is dynamically adjusted to the inverse frequency of positive examples in the training batch.

\subsubsection{Asynchronous Propagation Strategy}
Financial prediction tasks are characterized by a significant \textbf{Resolution Latency} ($\Delta T$). Standard methods often discard recent interactions within the interval $[t_{now} - \Delta T, t_{now}]$ due to the absence of ground-truth labels.

\begin{wrapfigure}{r}{0.5\columnwidth}
    \centering
    \includegraphics[width=\linewidth]{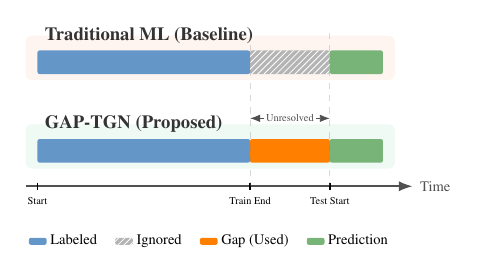}
    \caption{\textbf{Gap Utilization.} Standard models discard interaction data during the reporting gap.}
    \label{fig:gap_utilization}
\end{wrapfigure}

The \textbf{Asynchronous Propagation} strategy (detailed in Eq. 7) ensures that the memory state $\mathbf{s}_u(t)$ remains synchronized with the most recent entity activities, even when outcomes are latent.

This propagation signal $y_{\text{edge}}$ is integrated into the Graph Attention layer via $\mathbf{\phi}_j$. This allows the model to utilize the structural existence of recent trades for contextualization---for example, identifying a cluster of recent technology purchases---even while their performance outcomes remain pending. This ensures the node memory $\mathbf{s}_u(t)$ remains synchronized with the most recent entity activities. 
A visual comparison between our gap utilization strategy and traditional approaches is presented in Figure \ref{fig:gap_utilization}.
\vspace{-0.2cm}

\section{Experiments}

Our preliminary experimental evaluation focuses on establishing a between tabular baselines and the proposed GAP-TGN. We report results for the baseline model and the proposed model. We further outline the design for both Stage 2 and Stage 3 as a proposed extension to capture complex lobbying interactions, offering a roadmap for future research in political-financial graph networks.


\subsection{Experimental Setup}
\label{sec:experimental_setup}  

We evaluate GAP-TGN against three tabular baselines: Logistic Regression, Multi-Layer Perceptron (MLP), and XGBoost. These models utilize flattened feature vectors of all available attributes but ignore topological dependencies.

\subsubsection{Feature Sets}
\textbf{Trade Features ($\mathbf{x}_{\text{trade}}$):} We construct a 3-dimensional vector for each transaction containing: (1) log-transformed transaction amount in USD, (2) binary direction ($\mathbb{I}_{\text{buy}}$), and (3) log-transformed reporting lag in days.

\textbf{Market Features ($\mathbf{x}_{\text{mkt}}$):} To capture the systematic market environment, we engineer a 14-dimensional vector comprising:
\vspace{-0.2cm}
\begin{itemize}
    \item \textbf{Momentum:} Discrete cumulative returns over lookback windows of $\{1, 5, 10, 20\}$ days.
    \item \textbf{Volatility:} 20-day historical volatility ($v_{20d}$).
    \item \textbf{Technical Indicator:} 14-day Relative Strength Index (RSI) and Volume Ratio.
\end{itemize}
\vspace{-0.2cm}

\subsubsection{Target Generation}
Consistent with Equation~\ref{eq:target}, we set the threshold $\tau = 0\%$, labeling any trade that outperforms the S\&P 500 benchmark as the positive class ($y=1$).

To evaluate component contributions, we propose a roadmap distinguishing our current implementation from planned extensions. 
First, the \textbf{Transaction Baseline} serves as our present structural foundation, operating strictly on the transaction graph $\mathcal{G} = (\mathcal{V}, \mathcal{E}_{trade})$ with price context. 
As part of our immediate future work, \textbf{Node Enrichment} will augment this baseline with learnable embeddings for Legislator Party and State affiliations. 
Finally, \textbf{Edge Enhancement} is planned to introduce multi-relational context—including Lobbying, Donations, and Geo-Industrial ties—to capture soft influence channels alongside financial transactions.
All models are trained and tested on a monthly rolling basis from January 2019 to April 2024.

\begin{table}[htbp]
    \centering
    \begin{minipage}[t]{0.48\textwidth}
        \centering
        \caption{\textbf{Long-Horizon Performance}: GAP-TGN comparison at 18M/24M.}
        \label{tab:results_summary}
        \resizebox{\textwidth}{!}{
            \begin{tabular}{lcccc}
                \toprule
                & \multicolumn{2}{c}{\textbf{AUROC}} & \multicolumn{2}{c}{\textbf{F1-Score}} \\
                \textbf{Model} & 18M & 24M & 18M & 24M \\
                \midrule
                Logistic Reg. & 0.486 & 0.465 & 0.300 & 0.195 \\
                MLP & 0.496 & 0.498 & 0.402 & 0.378 \\
                XGBoost & \textbf{0.525} & \textbf{0.524} & 0.334 & 0.291 \\
                \textbf{GAP-TGN} & 0.490 & 0.518 & \textbf{0.438} & \textbf{0.440} \\
                \bottomrule
            \end{tabular}
        }
    \end{minipage}
    \hfill 
    \begin{minipage}[t]{0.48\textwidth}
        \centering
        \caption{\textbf{Ablation Study}: Component contribution analysis.}
        \label{tab:ablation_study}
        \resizebox{\textwidth}{!}{
            \begin{tabular}{lcccc}
                \toprule
                & \multicolumn{2}{c}{\textbf{AUROC}} & \multicolumn{2}{c}{\textbf{F1-Score}} \\
                \textbf{Model} & 18M & 24M & 18M & 24M \\
                \midrule
                \textbf{GAP-TGN (Full)} & \textbf{0.490} & 0.518 & \textbf{0.438} & \textbf{0.440} \\
                Vanilla TGN & 0.478 & 0.517 & 0.416 & 0.393 \\
                No Gated Fusion & 0.476 & 0.508 & 0.431 & 0.427 \\
                No Async Prop. & 0.478 & \textbf{0.521} & 0.424 & 0.405 \\
                \bottomrule
            \end{tabular}
        }
    \end{minipage}
    \vspace{-0.3cm}
\end{table}

\subsection{Preliminary Performance Analysis}

Table \ref{tab:results_summary} and Table \ref{tab:ablation_study} present the evaluation of our Transaction Baseline. 
A distinct trade-off emerges between ranking capability and classification stability. While tabular baselines like XGBoost achieve a marginally higher AUROC (0.524 at 24M) compared to GAP-TGN (0.518), they suffer from severe signal decay when forced to make binary actionable predictions. Specifically, XGBoost's F1-Score collapses to \textbf{0.291} at the 24-month horizon. 
In sharp contrast, GAP-TGN demonstrates superior \textbf{signal stability}, maintaining a robust F1-Score of \textbf{0.440}—an improvement of over \textbf{51\%} in identifying positive cases.

The ablation study further validates our architectural decisions. Removing the \textit{Asynchronous Propagation} strategy causes the F1-Score to drop to 0.405, confirming that bridging the unresolved gap is critical for long-horizon memory retention. Similarly, the \textit{Vanilla TGN} without market fusion degrades to 0.393, underscoring the necessity of contextualizing structural patterns with market data.
Overall, these preliminary results suggest that while the current structural baseline effectively captures persistent latent factors with high F1, it likely remains feature-starved relative to the tabular models with lower AUROC. 
We posit that the planned \textbf{Node Enrichment} and \textbf{Edge Enhancement} will provide the necessary informational density to close the AUROC gap while preserving the model's structural robustness.

\section{Conclusion}
In this work, we introduced \textit{Capitol Gains}, a dynamic graph dataset for modeling the legislative financial ecosystem. 
Preliminary results from our GAP-TGN \textbf{Transaction Baseline} suggest that structural signals provide superior classification stability compared to tabular baselines, even without auxiliary features. 
To support further exploration, we provide a detailed proposal for a \textbf{Downstream Financial Application} in Appendix~\ref{sec:application}, and a comprehensive review of \textbf{Related Work} in Appendix~\ref{app:related-work}, demonstrating the dataset's potential utility in realistic market settings. Future work will focus on completing the roadmap by integrating the planned node and edge enrichment modules.

\clearpage

\bibliography{references}
\bibliographystyle{iclr2026_conference}

\appendix

\section{Dataset Statistics}\label{app:statistics}

\begin{table}[h]
    \centering
    \caption{Dataset Statistics and Sizes}\label{tab:statistics}
    \label{tab:dataset_stats}
    \begin{tabular}{l l r}
        \hline
        \textbf{Dataset} & \textbf{Source} & \textbf{Entries} \\
        \hline
        Congressional Trades & Quiver Quantitative & 32,401 \\
        Legislator Metadata for Individual Terms & congress-legislators & 45,525 \\
        Roll Call Votes & VoteView & 26,262,296 \\
        Ideology Scores (Quarterly) & VoteView & 32,740 \\
        District Employment & American Community Survey & 5,000 \\
        Lobbying Reports & LobbyView & 1,690,362 \\
        PAC Transactions & OpenSecrets & 423,350,920 \\
        527 Org Transactions & OpenSecrets & 8,714,809 \\
        Corporate Financial Quarterly Indicators & SEC's EDGAR API & 20,309,853 \\
        Economic Indicators & FRED's ALFRED API & 260,778 \\
        \hline
    \end{tabular}
\end{table}

\section{Downstream Financial Application}\label{sec:application}


\begin{figure}[ht]
    \centering
    \includegraphics[width=\textwidth]{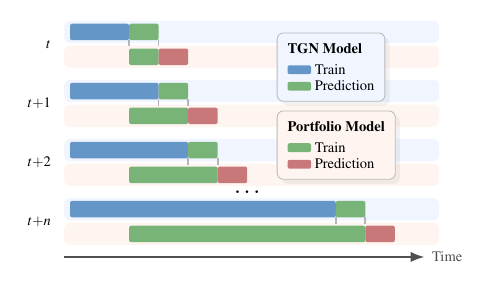}
    \caption{Two-level walk-forward training of GAP-TGN and portfolio allocation models.}
    \label{fig:walkforward}
\end{figure}

As this work represents an ongoing research, our current focus has been on establishing the quality of the \textit{Capitol Gains Dataset} and the architectural validity of GAP-TGN. 
To fully assess the economic utility of these structural signals in a realistic market setting, we propose a future backtesting environment based on a \textbf{Time-Series Stacking Architecture}.

\subsection{Portfolio Allocation Model}The proposed portfolio allocation model redetermines the portfolio configuration daily by processing prediction labels associated with each trade beginning on the day of the filing. Each label is interpreted as a confidence signal. The allocation model can be descirbed in four steps:

\begin{enumerate}
\item \textbf{Basket Confidence:} Measure the aggregate confidence in the active basket of stocks.
\item \textbf{Stock Confidence:} Measure confidence in each individual stock based on the signal strength from the TGN and apply the method of meta-labeling for bet sizing \citep{lopezdeprado2018advances}.
\item \textbf{Risk Adjustment:} Adjust weights according to risk tolerance and each stock's historical volatility via inverse volatility weighting to normalize risk contributions.

\item \textbf{Final Weighting:} Determine the final capital weights to assign to each stock and the S\&P 500 benchmark.\end{enumerate}

The model incorporates a signal decay mechanism. This concept of alpha decay ensures that older signals exercise decreasing influence on the current allocation \citep{grinold1999active}. The allocation function is governed by seven hyperparameters: S\&P-Stock weight balance, confidence sensitivity, volatility sensitivity, risk sensitivity, long-short weight, signal decay rate, and a threshold for signal activation. These parameters are optimized using a genetic algorithm \citep{brabazon2008evolutionary} which evolves the parameter set over the training window.

\subsection{Time-Series Stacking Architecture}We adopt the framework of stacked generalization \citep{wolpert1992stacked} adapted for temporal data structures as described by \citet{pardo2008evaluation}. As shown in Figure~\ref{fig:walkforward}, the architecture will consist of two levels. Level 0 is the TGN described in previous sections. It ingests the raw graph data and outputs a probability score for each congressional transaction. Level 1 is the portfolio allocation model. It ingests the out-of-sample probability scores from Level 0 as trade signals.

The training and testing windows for both models move in a coordinated walk-forward sequence. For any given testing month, the Level 0 model is trained on a rolling window of 24 months. The out-of-sample predictions generated by the TGN during the validation phase are concatenated to form the training set for the portfolio allocation model.

\subsection{Planned Comparative Metrics}
Upon completion of the financial backtest scheduled for the period from January 2016 to December 2025, we will compare the performance of five portfolios: a passive S\&P 500 benchmark, a portfolio driven by the Baseline (XGBoost) model, and three portfolios driven by the Stages 1-3 TGN models.We will assess the models using the following metrics:

\begin{itemize}

\item \textbf{Annualized Return and Volatility:} To measure raw performance and risk.

\item \textbf{Sharpe Ratio:} To measure the efficiency of returns per unit of risk.\item \textbf{Maximum Drawdown:} To assess the downside risk relative to the benchmark.

\item \textbf{Sortino and Calmar Ratios:} To provide further nuance regarding downside volatility and recovery potential.\end{itemize}We anticipate analyzing cumulative returns to observe divergence from the benchmark, specifically during periods of increased market volatility where the structural features of the TGN may provide a distinct informational advantage. Additionally, we will analyze the rolling Sharpe ratio to assess the temporal stability of risk-adjusted returns.

\section{Related Works}\label{app:related-work}

We classify the existing literature on congressional trading into three categories: historical performance relative to regulation, market-based predictors of outcomes, and performance attributed to specific leadership roles or regime trends. Table \ref{tab:related_works} summarizes these findings.

\begin{table}[ht]
    \centering
    \caption{Classification of Related Works on Congressional Trading}
    \label{tab:related_works}
    \begin{tabular}{p{0.2\textwidth} p{0.3\textwidth} p{0.45\textwidth}}
        \hline
        \textbf{Category} & \textbf{Study} & \textbf{Key Findings} \\
        \hline
        Historical \& Regulation 
        & \citet{ziobrowski2004abnormal} 
        & Documented significant informational advantage with $\sim$12\% annual abnormal returns in the 1990s. \\
        
        & \citet{eggers2013capitol} 
        & Corrected datasets show congressional portfolios underperformed passive benchmarks (2004--2008). \\
        
        & \citet{belmont2022do}; \citet{karadas2021did} 
        & Confirmed market efficiency post-STOCK Act (2012); Act reduced predictive power regarding macro information. \\
        \hline
        Predictors \& Instruments 
        & \citet{karadas2024what}; \citet{vo2021does} 
        & Trading volume correlates positively with Economic Policy Uncertainty (EPU); aggressive trading occurs when public data is unreliable. \\
        
        & \citet{molk2025negative} 
        & Identifies performance asymmetry; short positions generate significant abnormal returns while long positions do not. \\
        \hline
        Roles \& Sectors 
        & \citet{wei2025captain} 
        & Leadership positions realize $\sim$47\% annualized abnormal returns; rank-and-file members exhibit standard performance. \\
        
        & \citet{kuziemko2025congressional} 
        & Public exposure to high-performing cohorts reduces trust in government and compliance with federal law. \\
        
        & \citet{motley2026congressional} 
        & In 2024, Democrats outperformed Republicans (31--33\% vs 26\%) due to heavier allocation in the technology sector. \\
        
        & \citet{truthout2024defense}; \citet{responsible2024defense} 
        & Defense committee members traded defense contractors during periods of geopolitical tension. \\
        \hline
    \end{tabular}
\end{table}


\section{Macroeconomic Indicators}
\label{app:macro_indicators}

We collect 64 macroeconomic indicators to serve as global context features. These are updated daily based on the most recent value reported by the Federal Reserve Bank of St. Louis (FRED/ALFRED) at that time.

\begin{longtable}{p{0.35\textwidth} p{0.6\textwidth}}
\caption{List of Macroeconomic Indicators} \label{tab:macro_indicators} \\
\hline
\textbf{Category} & \textbf{Indicators (Ticker)} \\
\hline
\endfirsthead
\hline
\textbf{Category} & \textbf{Indicators (Ticker)} \\
\hline
\endhead
\hline
\endfoot

Prices \& Inflation & Consumer Price Index (CPIAUCSL), Core CPI (CPILFESL), PCE Price Index (PCEPI), Core PCE Price Index (PCEPILFE), Producer Price Index (PPIACO), GDP Price Deflator (GDPDEF) \\
\hline
Monetary Policy \& Interest Rates & Federal Funds Rate (FEDFUNDS), 10-Year Treasury Rate (DGS10), 2-Year Treasury Rate (DGS2), M2 Money Supply (M2SL), 10-Year TIPS (DFII10) \\
\hline
Economic Output \& Activity & Real GDP (GDPC1), Potential GDP (GDPPOT), Industrial Production (INDPRO), Housing Starts (HOUST) \\
\hline
Labor Market & Unemployment Rate (UNRATE), Nonfarm Payrolls (PAYEMS), Initial Claims (ICSA), Job Openings (JTSJOL) \\
\hline
Commodities \& Raw Materials & WTI Crude Oil Price (DCOILWTICO), Brent Crude Oil Price (DCOILBRENTEU), U.S. Regular Gas Price (GASREGCOVW) \\
\hline
Market Sentiment \& Expectations & CBOE Volatility Index (VIXCLS), Consumer Sentiment (UMCSENT) \\
\end{longtable}

\section{Corporate Financial Facts}
\label{app:corporate_facts}

The following financial facts are extracted from quarterly 10-Q filings via the SEC EDGAR API. These features are indexed by their filing date to prevent lookahead bias.

\begin{itemize}
    \item Accounts Payable Current
    \item Accounts Receivable Net Current
    \item Accumulated Other Comprehensive Income Loss Net Of Tax
    \item Additional Paid In Capital
    \item Amortization Of Intangible Assets
    \item Assets
    \item Assets Current
    \item Cash And Cash Equivalents At Carrying Value
    \item Common Stock Dividends Per Share Declared
    \item Common Stock Value
    \item Earnings Per Share Basic
    \item Earnings Per Share Diluted
    \item Goodwill
    \item Gross Profit
    \item Income Tax Expense Benefit
    \item Interest Expense
    \item Liabilities
    \item Liabilities And Stockholders Equity
    \item Liabilities Current
    \item Net Cash Provided By Used In Financing Activities
    \item Net Cash Provided By Used In Investing Activities
    \item Net Cash Provided By Used In Operating Activities
    \item Net Income Loss
    \item Nonoperating Income Expense
    \item Operating Expenses
    \item Operating Income Loss
    \item Other Assets Noncurrent
    \item Payments To Acquire Property Plant And Equipment
    \item Property Plant And Equipment Net
    \item Research And Development Expense
    \item Retained Earnings Accumulated Deficit
    \item Revenues
    \item Selling General And Administrative Expense
    \item Stockholders Equity
    \item Weighted Average Number Of Diluted Shares Outstanding
    \item Weighted Average Number Of Shares Outstanding Basic
\end{itemize}



\end{document}